\documentclass[%
reprint,
amsmath,amssymb,
aps,
]{revtex4-2}

\usepackage{graphicx}
\usepackage{dcolumn}
\usepackage{bm}
\usepackage{subfig}

\usepackage{tikz}
\usetikzlibrary{arrows, shapes, chains}
\tikzstyle{startstop} = [rectangle,rounded corners, minimum width=3cm,minimum height=1cm,text centered,text width=8.5cm, draw=black,fill=gray!30]
\tikzstyle{io} = [trapezium, trapezium left angle = 70,trapezium right angle=110,minimum width=3cm,minimum height=1cm,text centered,text width=9cm,draw=black,fill=gray!30]
\tikzstyle{process} = [rectangle,minimum width=3cm,minimum height=1cm,text centered,text width =3cm,draw=black,fill=gray!30]
\tikzstyle{decision} = [diamond,minimum width=3cm,minimum height=1cm,text centered,draw=black,fill=gray!30]
\tikzstyle{arrow} = [thick,->,>=stealth]

\begin{document}


\title{An explanation for the distribution characteristics of stock returns}


\author{Bo Li}
\email[]{libo312@mails.ucas.ac.cn}
\affiliation{PKU-WUHAN Institute for Artificial Intelligence}


\date{\today}

\begin{abstract}
Observations indicate that the distributions of stock returns in financial markets usually do not conform to normal distributions, but rather exhibit characteristics of high peaks, fat tails and biases.
In this work, we assume that the effects of events or information on prices obey normal distribution, while financial markets often overreact or underreact to events or information, resulting in non normal distributions of stock returns.
Based on the above assumptions, we propose a reaction function for a financial market reacting to events or information, and a model based on it to describe the distribution of real stock returns.
Our analysis of the returns of China Securities Index 300 (CSI 300), the Standard \& Poor’s 500 Index (SPX or S\&P 500) and the Nikkei 225 Index (N225) at different time scales shows that financial markets often underreact to events or information with minor impacts, overreact to events or information with relatively significant impacts, and react slightly stronger to positive events or information than to negative ones. 
In addition, differences in financial markets and time scales of returns can also affect the shapes of the reaction functions.
\end{abstract}


\maketitle

\section{Introduction} \label{sec_intro}
The distribution of stock returns plays a fundamental role in financial research and practice, which is crucial for risk management, asset pricing, portfolio management, and etc.
The distribution of stock returns is generally treated as normal distribution or lognormal distribution, which was first proposed by Kendall and Osborne \cite{kendall1953analysis, osborne1959brownian}.
In the standard financial theory system, Black and Sholes' Option Pricing Model is based on stock returns follow a lognormal distribution \cite{black1973pricing},
the Capital Asset Pricing Model (CAPM) assumes that the returns distribution of each security follows a normal distribution and the joint distribution of returns of different securities obeys multivariate normal distribution \cite{sharpe1964capital}.
J. P. Morgan's VAR system risk metric is also based on the assumption that the return rates of securities follow normal distributions \cite{phelan1997probability}.

Intuitively, from the mathematical point of view, stock returns satisfy that the sample size is large enough and they are influenced by multiple different factors, which are the prerequisites for the central limit theorem \cite{fischer2011history}. 
Therefore, the idea that stock returns follow normal distribution looks very natural.
In addition, adopting the normal distribution means that dealing with related financial problems in research and application is easy.
However, although normal distribution is widely used when describing distribution of stock returns, it can not characterize the characteristics of high peaks and fat tails displayed by the true stock returns distribution.
Many analyses of asset returns indicate that the distribution of price returns has the characteristics of high peaks, fat tails, and skewnesses \cite{alexander1961price, peters1991chaotic, wang2009statistical}.
Besides, researches on returns series have shown that there are linear and nonlinear correlations in returns series \cite{liu1999statistical, bogachev2007effect, wang2011quantifying, shi2021clustering}.

Since normal distributions could not characterize the distributions of stock returns well, many researchers have conducted further studies on this issue, attempting to find appropriate distribution models with taking into account high peak, fat tail, and skewness characteristics of stock returns.
The distributions proposed by researchers include:
logistic distribution, which is similar to normal distribution but has a fatter tail \cite{smith1981probability, gray1990empirical, peiro1994distribution};
exponential power distribution, which has the characteristics of high peak and thick tail, and the tail shrinks exponentially, thus providing a good fit for the stock returns distribution \cite{hsu1982bayesian, gray1990empirical};
Possion mixture of normal distributions, which believes that the stock price fluctuations are composed of a continuous diffusion (Brownian motion) and a discontinuous jump (Poisson point process), the former causes continuous changes in stock prices, while the latter reflects the significant shock brought by the material news \cite{press1967compound};
mixed normal distribution, Kon provided some empirical evidence for this mixed normal distribution \cite{kon1984models};
scaled-t distribution, which assumes that the volatility of stock returns is a time-varying random variable, and can better fit the returns distribution than normal distribution \cite{praetz1972distribution, blattberg1972comparison, gray1990empirical, aparicio2001empirical};
truncated l\'{e}vy distribution, which was found to fit the distribution of the S\&P 500 index well  \cite{mantegna1994stochastic, mantegna1995scaling, romanovsky2000truncated}.

Scholars have also conducted researches on the mechanisms why stock returns do not conform to normal distributions. 
Peters et al. \cite{peters1996chaos} explored this issue from the perspective of information dissemination and believed that information is transmitted to the market in a clustered manner, rather than arriving timely in a linear manner, resulting in a peak distribution of information and thus affecting the distribution of stock returns.
Based on behavioral finance, some researchers believe that it is caused by momentum effects, herd effects, and other behavior factors in the financial market \cite{de2012exchange, kothari2006stock}.
Some scholars believe that volatility is time-varying, and explain the high peak and fat tail shape of the returns distribution based on it \cite{kaur2004time, zhang2010reexamination}.

This paper argues that the effects of various events or information on the financial market is normally distributed.
However, due to people's overreaction or underreaction to events or information, the stock returns distribution deviates from the normal distribution and exhibits the characteristics of high peaks, heavy tails and skewnesses.
The above hypothesis has theoretical supports from behavioral finance.
In behavioral finance, the overreaction theory based on overreaction behaviors of investors is one of the four major research achievements in finance, which are the overreaction theory, the prospect theory, the regret theory, and the overconfidence theory \cite{howe1986evidence, barberis2003survey, baker2010behavioral}. 
Based on the assumption, this article proposes a reaction function of the financial market reacting to various types of information or events, provides method for calculating the reaction function curve using real stock returns and utilize the calculation results to explain the distribution of asset returns in the financial market.

The innovation and main contribution of this article lies in explaining the distribution characteristics of real stock returns from a new perspective, proposing the reaction function of financial markets to events or information, and providing the calculation process of this reaction function through real stock returns.
It must be pointed out that the reaction function in this study represents the average effect, which is calculated based on the difference between the hypothetical normal distribution and the true distribution of a large number of returns, and the distribution of returns is a statistical result formed by a large number of samples.

The structure of this paper is as follows. 
We propose our model in Sec.~\ref{sec_model}. 
In Sec.~\ref{sec_exp}, we introduced the data used in the experiment, the calculation process, and the analysis and discussion of the experimental results. 
Sec.~\ref{sec_con} is the conclusion.

\section{Model}\label{sec_model}
The rate of return on assets in financial markets is influenced by a combination of multiple factors or events.
We assume that, in terms of their impacts on the assets returns, the combined impacts of these factors or events follow a normal distribution.
On this basis, we assume that the reason why the rate of return on assets exhibits the distribution characteristics of peak and thick tail is because the reactions of the financial markets to the comprehensive impacts of these events or factors are distorted, existing overreaction or underreaction.
Under our assumption, if the financial market's reactions to the combination of multiple factors or events are not distorted, the distribution of the returns should be normal and can be denoted as $N(\mu,\sigma^2)$.
In the following text, we refer to it as the hypothetical raw normal distribution.

We make the above assumption based on two reasons. 
The first one is related to the central limit theorem, which states that the sample size is large enough, the sampling distribution of the mean of the variable will approximate a normal distribution, independent of the distribution of the variable in the population.
For the financial markets, the returns are influenced by multiple factors, and the degrees of these comprehensive factors' influences on assets' prices should follow a normal distribution.
The other reason is from the perspective of behavioral finance, the rate of return exhibits the distribution characteristics of peak and thick tail may be due to people's overreaction or underreaction to information that affects the financial market.

Based on the above, we denote the probability density function of asset price returns in the real market as $f_t\left (r^{true}\right )$, and there is the following relationship
\begin{equation}
	r^{true}-\mu^{true} = R(r) (r-\mu) \label{eq1}
\end{equation}
where $r\in N(\mu,\sigma^2)$ and $R(r)$ is the reaction function, which reflects the market's reaction intensity of events or information, $\mu$ and $\mu^{true}$ represent the hypothetical and real mean value of returns, respectively. 
If $R(r)=1$, there is no distortion in the market's responses to events or information. 
If $R(r)<1$, it indicates insufficient market responses to information or events. 
If $R(r)>1$, it indicates excessive market responses to events or information.
Based on behavioral finance and the actual distribution of returns, $R(r)$ should be a finite value.

We assume $r^{true}=\mu$ and define $R(\mu)=0$ when $r$ approaches $\mu$. 
The reason for making the above assumption is as follows: when $r=\mu$, that is, when the impact of events or information on stock prices is in the mean position, the market does not respond to events or information, and then there is $r^{true}=\mu^{true}$. 
Since $R(r)(r-\mu)=r^{true}-\mu^{true}=0$ at this point $r=\mu$ then $R(\mu)$ can take any value, the reason why we choose $R(\mu)=0$ is to make the reaction function curves calculated in the following text smoother.
Based on the above analysis, the true distribution density function can be expressed as
\begin{equation}
	f_t(r^{true}) = \left\{
	\begin{aligned}
		&\frac{1}{\sqrt{2\pi}\sigma}e^{-\frac{(r^{true}-\mu^{true})^2}{2\sigma^2 R^2_t(r^{true})}}\  if\  \mu \neq \mu^{true}, \\
		&\frac{1}{\sqrt{2\pi}\sigma}e^{-\frac{(r^{true}-\mu^{true})^2}{2\sigma^2 }}\  if\  \mu = \mu^{true},
	\end{aligned}
	\right.
\end{equation}
where $R_t(r^{true})$ is the value of the reaction function $R(r)$ when the true rate of return is $r^{true}$.
Therefore, the probability density value of the real asset price returns at the center point is consistent with the probability density value of $N(\mu,\sigma^2)$ at the center point. 

In real data processing, we calculate the 50\% percentile value in true returns to determine the value of $\mu$ and $\mu^{true}$.
For the value of $\sigma$, we can determine it in the following way.
Firstly, at a position close to the mean of the real returns, the probability distribution function of the real returns is consistent with the hypothetical raw normal distribution function. 
Therefore, we have
\begin{equation}
	F_t(\mu^{+})-F_t(\mu^{-}) \approx \frac{1}{\sqrt{2\pi }\sigma}(\mu^+-\mu^-),
\end{equation}
if $(\mu^+-\mu) /\sigma \to 0$ and $(\mu-\mu^-) /\sigma \to 0$, where $\mu^{+}$ and $\mu^{-}$ are values slightly greater than $\mu$ and slightly less than $\mu$ respectively, and $F_t(r^{true})$ is the cumulative distribution function of the true returns.
The cumulative distribution function of the true function $F_t(r^{true})$ is known, $\mu^+$ and $\mu^-$ can be selected as required, so
\begin{equation}
	\sigma \approx \frac{\mu^+-\mu^-}{\sqrt{2\pi}[F_t(\mu^{+})-F_t(\mu^{-})]}.  \label{eq_sigma}
\end{equation}

Assuming that the value of $R_t(r^{true})$ remains unchanged in a very small region $(r_1^{true}, r_2^{true})$, we can approximately calculate the value of $R_t(r^{true})$ for each interval $(r_1^{true}, r_2^{true})$.
First of all, according to the concept of cumulative distribution function we have
\begin{equation}
	F_t(r^{true}_2)-F_t(r^{true}_1) = \int_{r^{true}_1}^{r^{true}_2} f_t(r^{true}) d r^{true},
\end{equation}
where $r_1^{true} = R(r_1)r_1$ and $r_2^{true} = R(r_2)r_2$, and $F(r^{true}_2)-F(r^{true}_1)$ can be obtained from the true returns.
When $r_1^{true}$ and $r_2^{true}$ are very close, we assume that the probability distribution density function of the true returns $f_t(r^{true})$ keeps constant, so
\begin{equation}
\begin{aligned}
	\Delta F_t &= F_t(r^{true}_2)-F_t(r^{true}_1) \\
	&\approx f_t(r_0^{true}) (r^{true}_2 - r^{true}_1)\\
	&=\frac{r^{true}_2 - r^{true}_1}{\sqrt{2\pi}\sigma}e^{-\frac{(r_0^{true}-\mu^{true})^2}{2\sigma^2 R^2_t(r^{true})}},
\end{aligned}
\end{equation}
where $r_0^{true} = (r^{true}_1+r^{true}_2)/2$.
Then we can approximately solve for the value of $R_t(r^{true})$, which is 
\begin{equation}
	R_t(r^{true}) \approx \left\{
	\begin{aligned}
		&\frac{r_0^{true}-\mu^{true}}{\sqrt{-2\ln\frac{\sqrt{2\pi}\sigma\Delta F_t}{r^{true}_2 - r^{true}_1}}}\  if \  r^{true}_1 > \mu^{true},\\
		&\frac{-\left(r_0^{true}-\mu^{true}\right)}{\sqrt{-2\ln\frac{\sqrt{2\pi}\sigma\Delta F_t}{r^{true}_2 - r^{true}_1}}}\  if \  r^{true}_2 \leq \mu^{true} .\\
	\end{aligned} 
	\right. \label{eq_reaction}
\end{equation} 

\section{Experiments and analysis}\label{sec_exp}
\subsection{Calculation process and data}
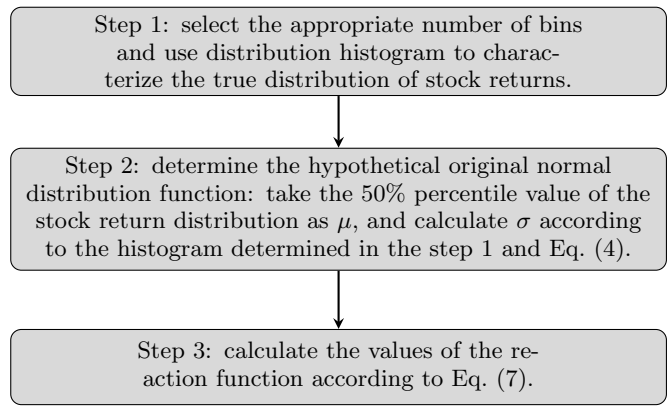
\begin{figure}[!h]
	\begin{center}
		\begin{tikzpicture}[node distance=1.5cm]
			\node (step1) [startstop] {Step 1: select the appropriate number of bins and use distribution histogram to characterize the true distribution of stock returns.};
			\node (step2) [startstop,below of=step1,yshift=-0.6cm] {Step 2: determine the hypothetical original normal distribution function: take the 50\% percentile value of the stock return distribution as $\mu$, and calculate $\sigma$ according to the histogram determined in the step 1 and Eq.~(\ref{eq_sigma}).};
			\node (step3) [startstop,below of=step2,yshift=-0.6cm] {Step 3: calculate the values of the reaction function according to Eq.~(\ref{eq_reaction}).};
			
			\draw [arrow] (step1) -- (step2);
			\draw [arrow] (step2) -- (step3);

		\end{tikzpicture}
	\end{center}
	\caption{The flow sketchmap of calculating the reaction function through the true stock returns.} \label{flow_skep}
\end{figure}

Fig.~\ref{flow_skep} shows the process of calculating the reaction function through the true returns in this study.
The first step is to select the appropriate number of bins and use distribution histograms to characterize the true distributions of stock returns.
Secondly, determine the hypothetical raw normal distribution function: take the 50\% percentile value of the stock returns distribution as $\mu$, and calculate $\sigma$ according to the histogram determined in the step 1 and Eq.~(\ref{eq_sigma}).
The last step is to calculate the values of the reaction function $R_t(r^{true})$ according to Eq.~(\ref{eq_reaction}).

We selected three representative stock price indexes in the world, China Securities Index 300 (CSI 300) from China, Standard \& Poor's 500 Index (SPX) from US, and Nikkei 225 Index (N225) from Japan.
For the daily market data of these indexes, CSI 300 is ranging from January 4, 2005 to November 1, 2023, SPX is ranging from September 20, 1982 to October 6, 2023, and N225 is ranging from March 25, 1985 to October 6, 2023.
For the minute level market data, CSI 300 is ranging from January 4, 2011 to February 2020, SPX is ranging from January 2, 2015 to July 6, 2020, and N225 is ranging from January 5, 2015 to July 7, 2020.
According to the process shown in Fig.~\ref{flow_skep}, we calculate the reaction function $R_t(r^{true})$ of the above indexes at daily, 15-minute, 5-minute, and 1-minute time scales in this work. 
The 1-minute time scale here refers to calculating the returns every minute, the meanings of the 5-minute and 15-minute time scales are similar.
When calculating the returns at minute level, we only count the results that occurred within the same trading day.

\subsection{Determine the appropriate number of bins}
We use histograms to characterize the probability distribution of true stock returns, which is the key step to determine the hypothetical raw normal distribution. 
When using histograms, the number of bins is crucial as it directly relates to the value of $\sigma$ in the hypothetical raw normal distribution density function. 

\begin{figure}[!htbp]
	\centering
	\subfloat[The true distribution histogram and the hypothetical raw normal distribution density function curve of CSI 300 daily returns when the number of bins is 100. The parameters of the hypothetical raw normal distribution are $\mu=6.5\times10^{-4}$ (keep two significant digits, the same below) and $\sigma=0.010$ (keep to three decimal places, the same below).]{
		\includegraphics[scale=0.15]{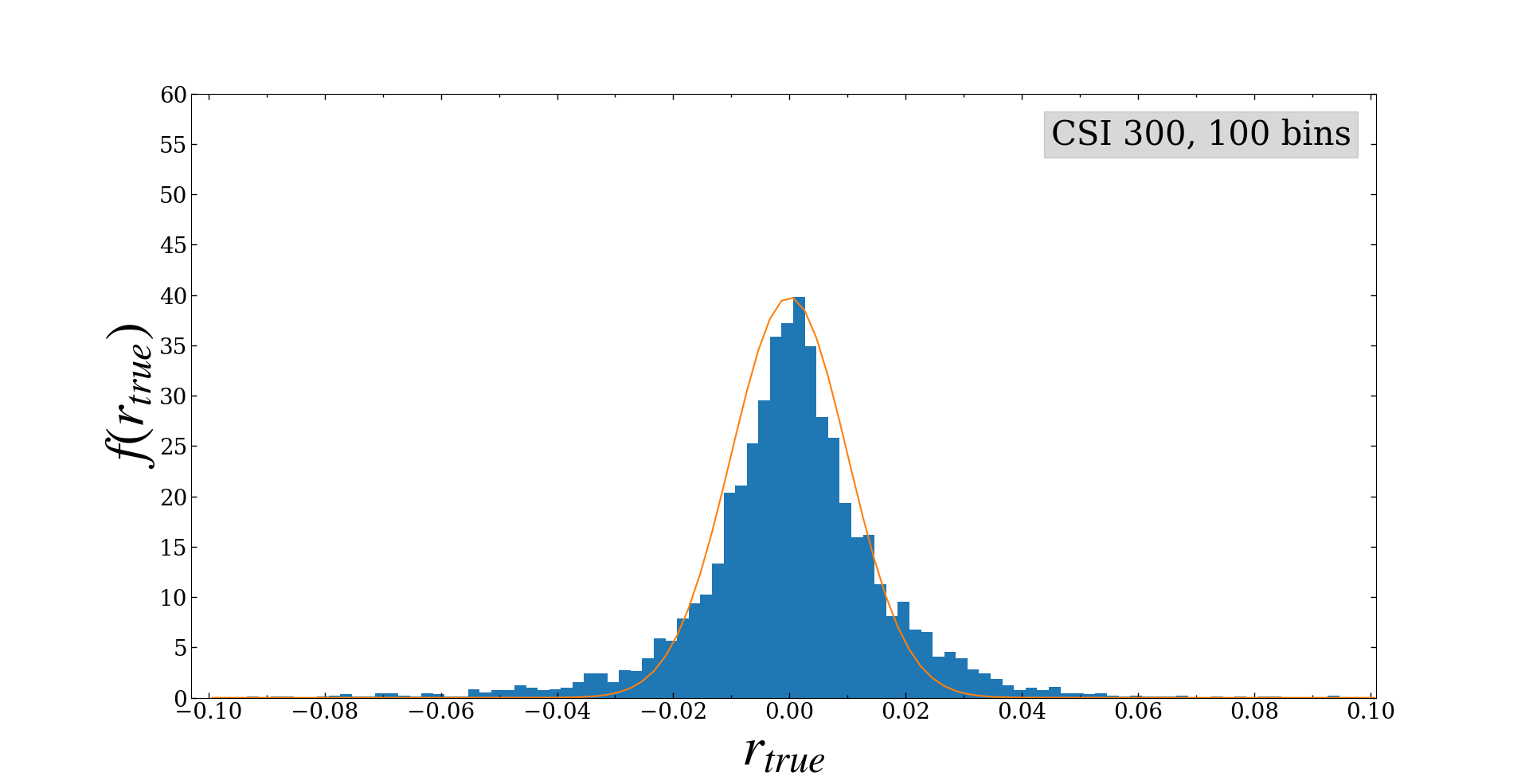}}\hspace{0mm}\\
	\subfloat[The true distribution histogram and the hypothetical raw normal distribution density function curve of CSI 300 daily returns when the number of bins is 150. The parameters of the hypothetical raw normal distribution are $\mu=6.5\times10^{-4}$ and $\sigma=0.009$.]{
		\includegraphics[scale=0.15]{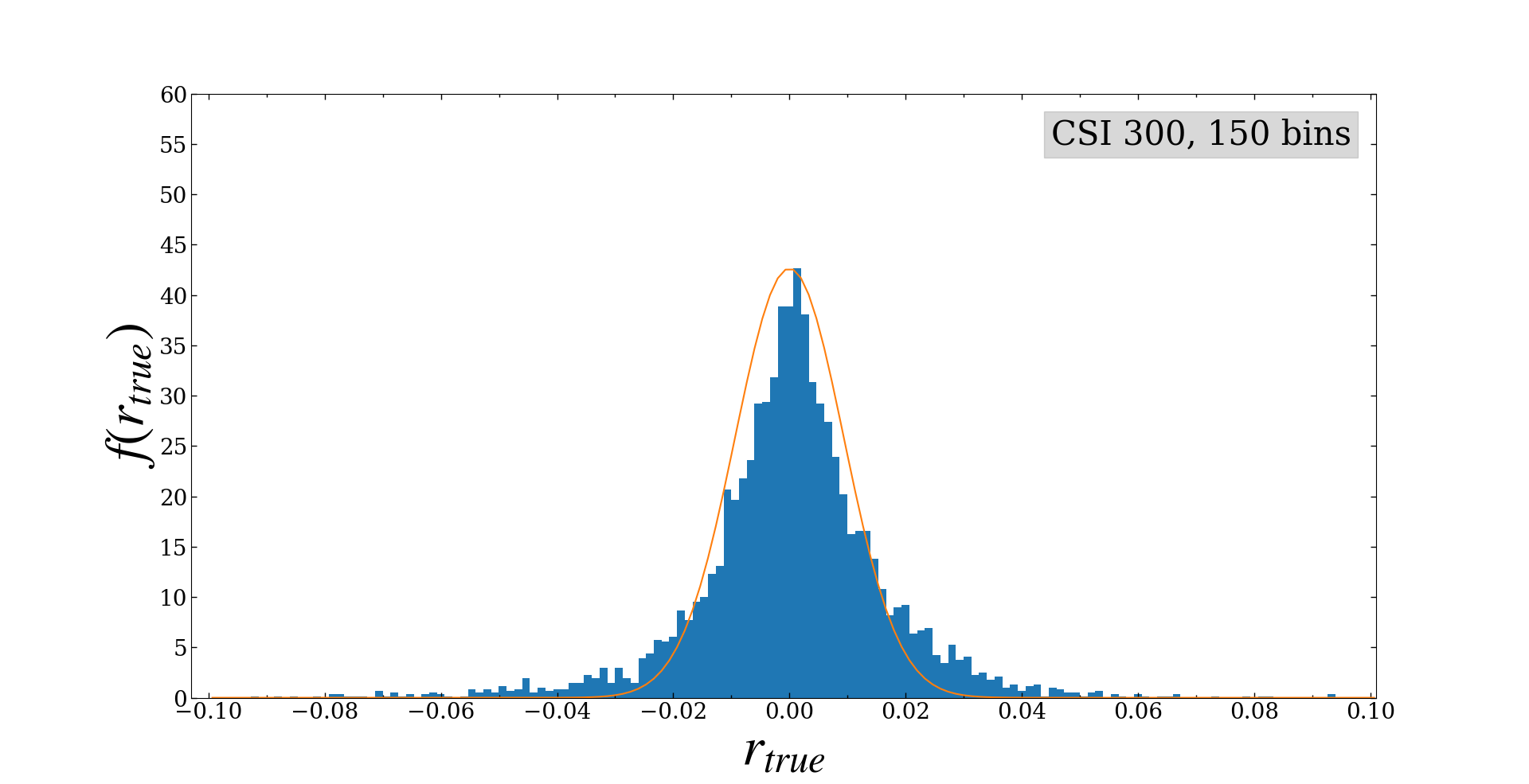}}\hspace{0mm}\\
	\subfloat[The true distribution histogram and the hypothetical raw normal distribution density function curve of CSI 300 daily returns when the number of bins is 200. The parameters of the hypothetical raw normal distribution are $\mu=6.5\times10^{-4}$ and $\sigma=0.010$.]{
		\includegraphics[scale=0.15]{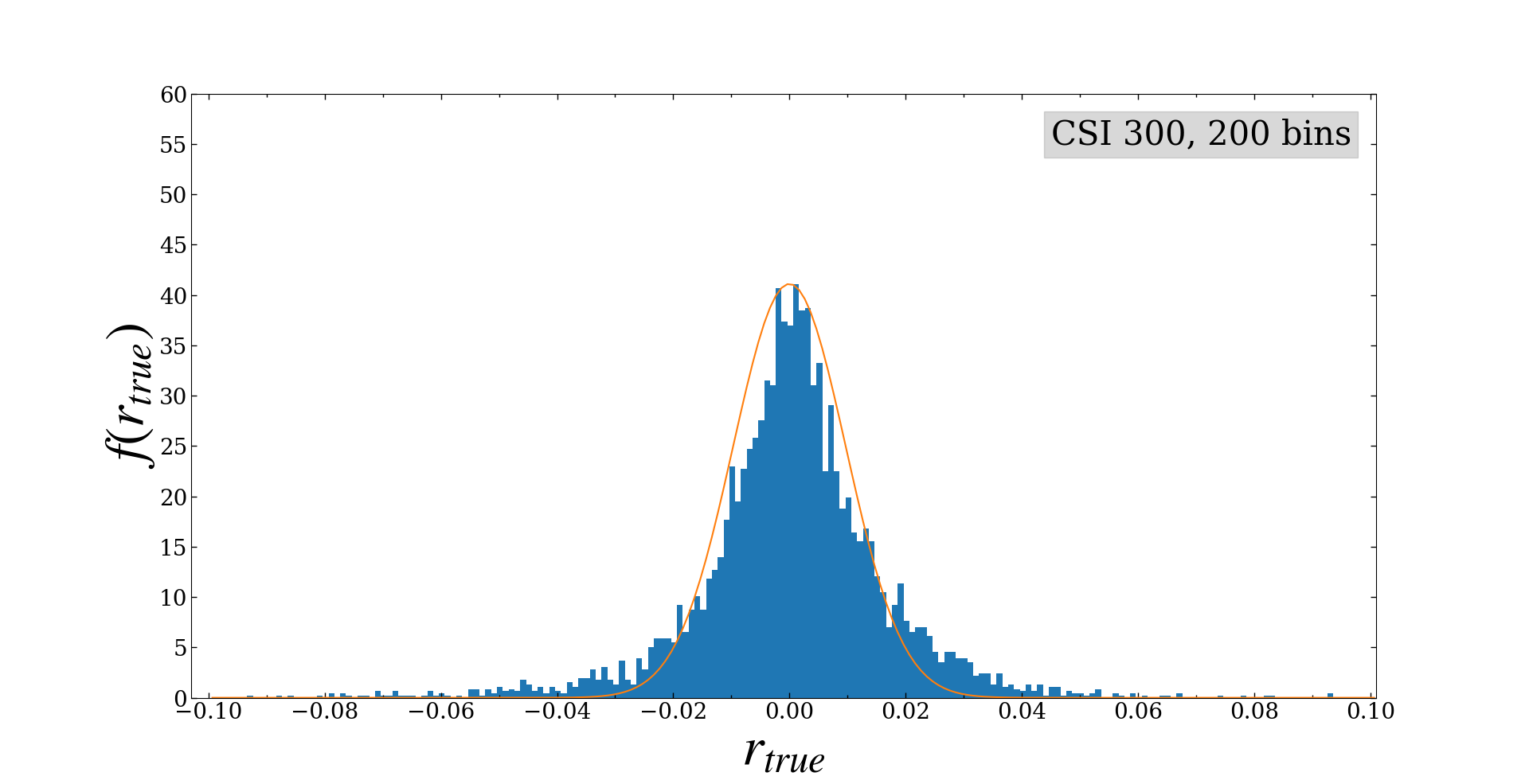}}
	\caption{The true distribution histograms and the hypothetical raw normal distribution density function curves of CSI 300 daily returns when taking different numbers of bins.}
	\label{CSI_d}
\end{figure}
\begin{figure}[!htbp]
	\centering
	\subfloat[The true distribution histogram and the hypothetical raw normal distribution density function curve of SPX daily returns when the number of bins is 100. The parameters of the hypothetical raw normal distribution are $\mu=6.0\times10^{-4}$ and $\sigma=0.007$.]{
		\includegraphics[scale=0.15]{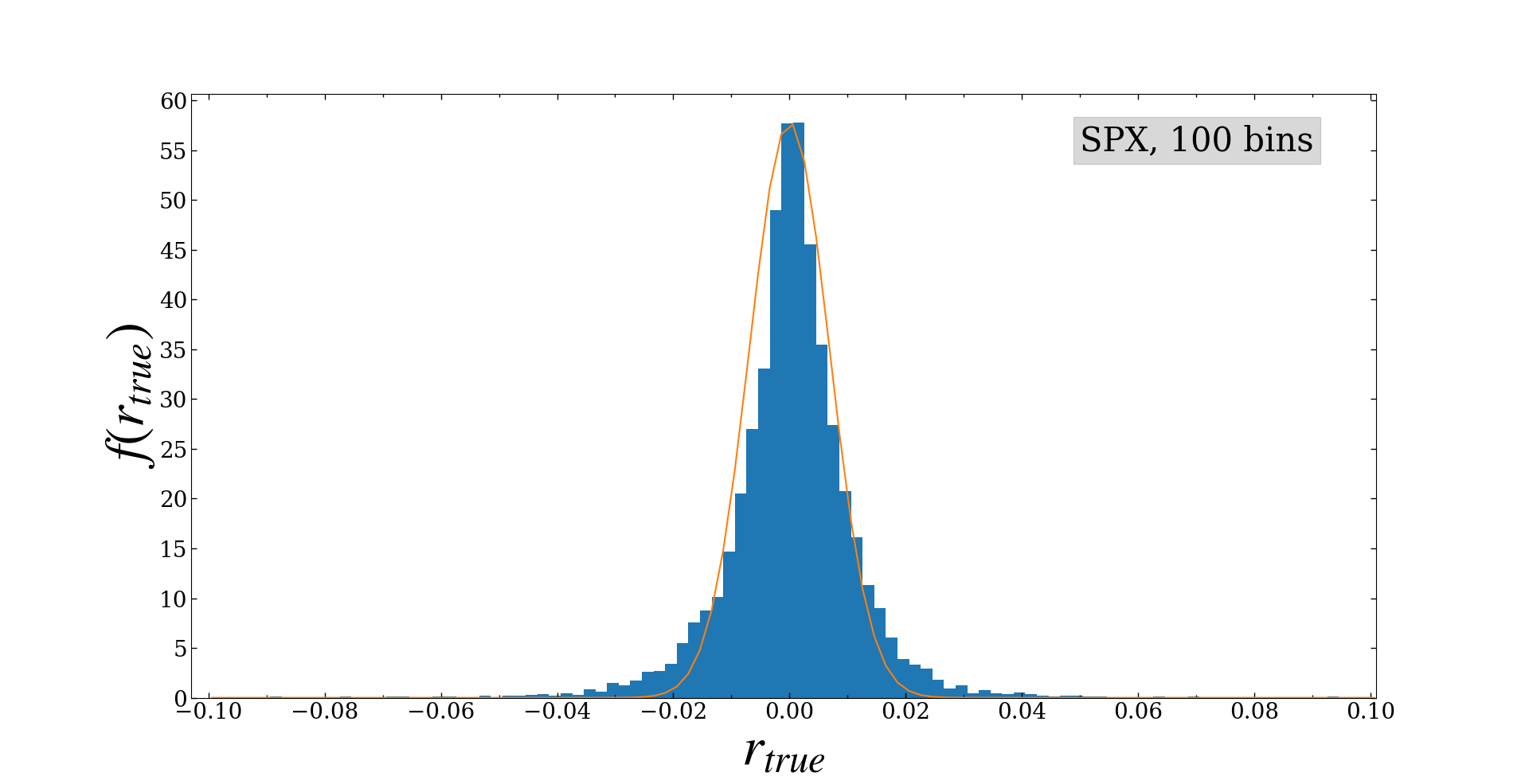}}\hspace{0mm}\\
	\subfloat[The true distribution histogram and the hypothetical raw normal distribution density function curve of SPX daily returns when the number of bins is 150. The parameters of the hypothetical raw normal distribution are $\mu=6.5\times10^{-4}$ and $\sigma=0.007$.]{
		\includegraphics[scale=0.15]{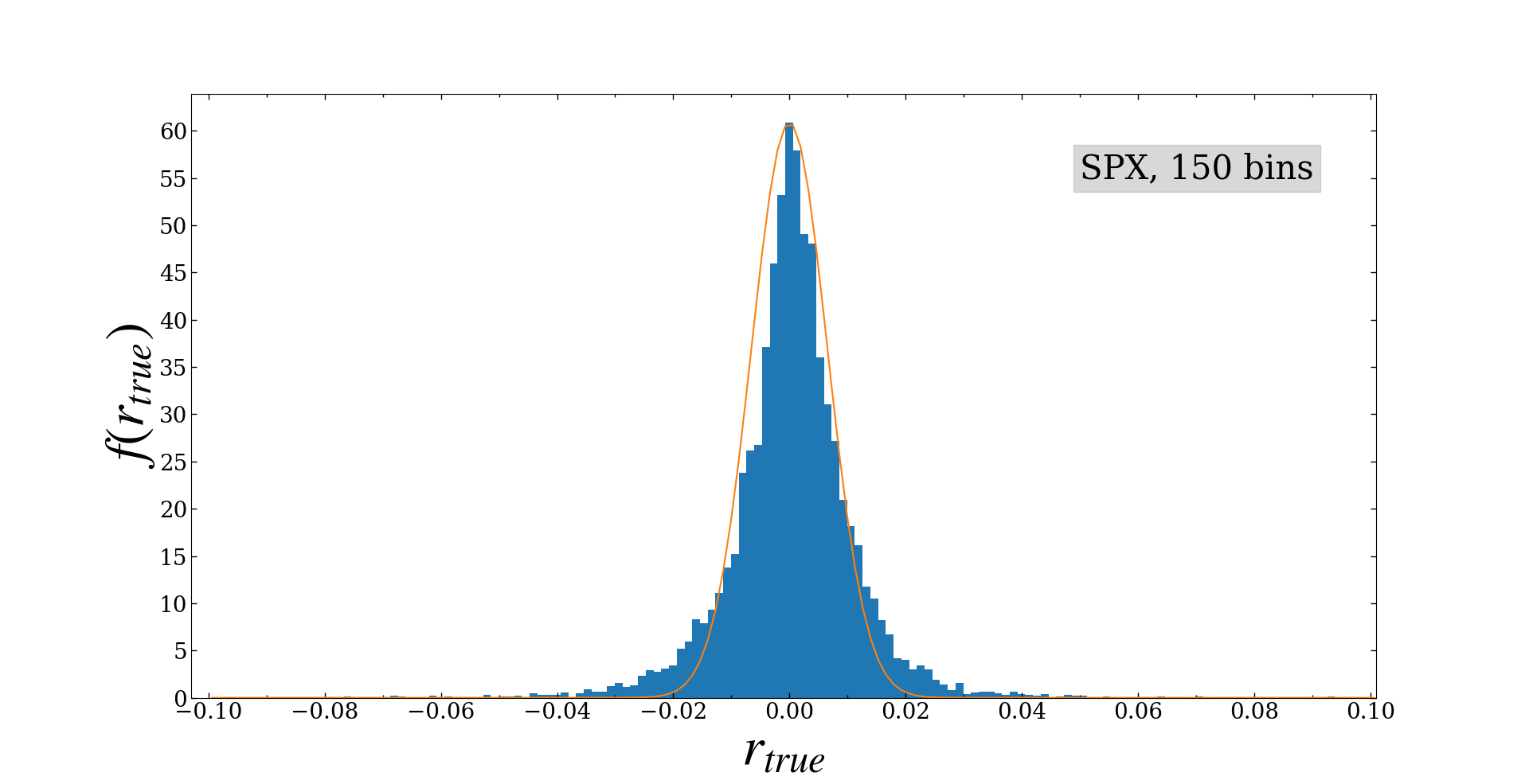}}\hspace{0mm}\\
	\subfloat[The true distribution histogram and the hypothetical raw normal distribution density function curve of SPX daily returns when the number of bins is 200. The parameters of the hypothetical raw normal distribution are $\mu=6.0\times10^{-4}$ and $\sigma=0.007$.]{
		\includegraphics[scale=0.15]{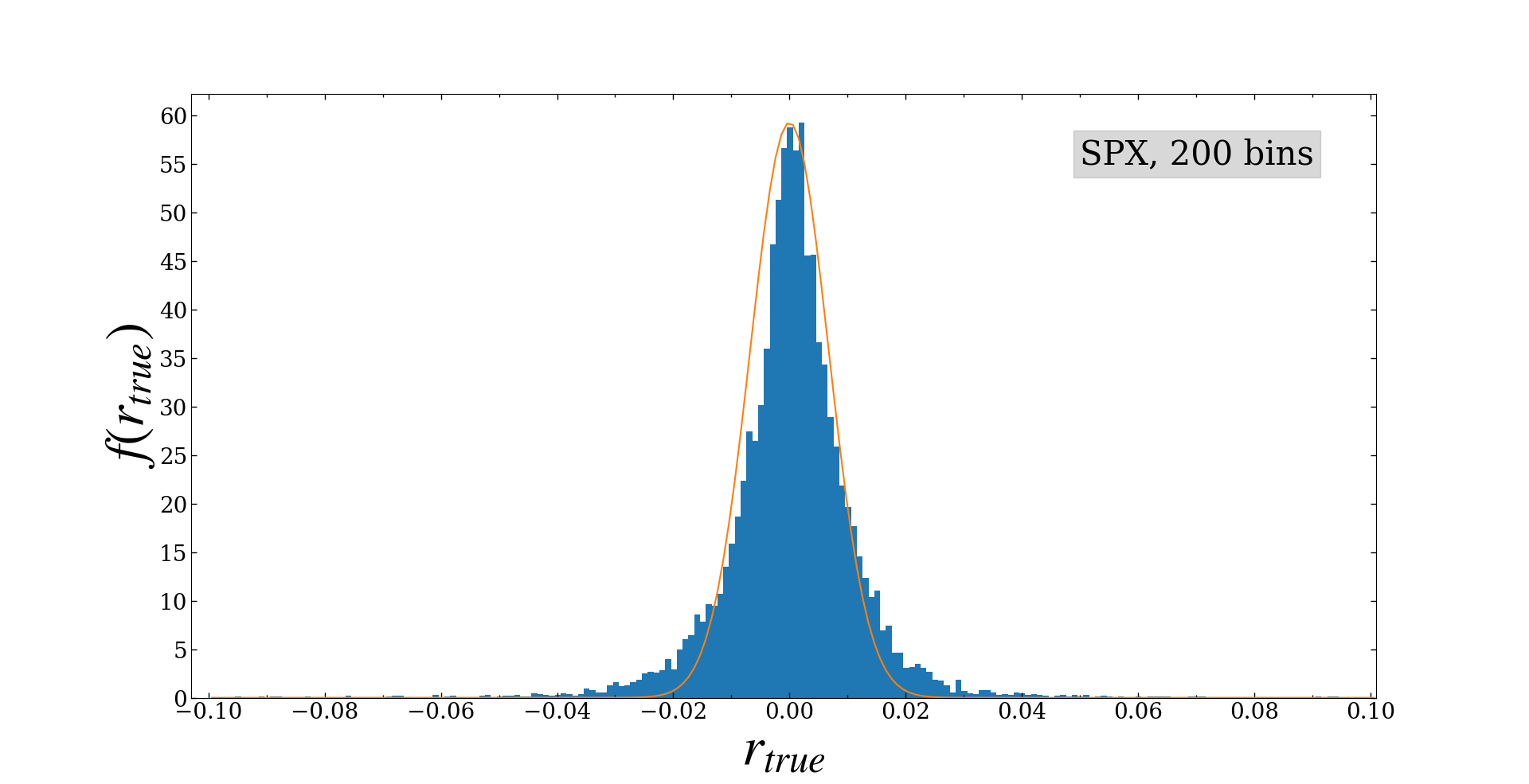}}
	\caption{The true distribution histograms and the hypothetical raw normal distribution density function curves of SPX daily returns when taking different numbers of bins.}
	\label{SPX_d}
\end{figure}
\begin{figure}[!htbp]
	\centering
	\subfloat[The true distribution histogram and raw normal distribution density function curve of N225 daily returns when the number of bins is 100. The parameters of the hypothetical raw normal distribution are $\mu=4.0\times10^{-4}$ and $\sigma=0.010$.]{
		\includegraphics[scale=0.15]{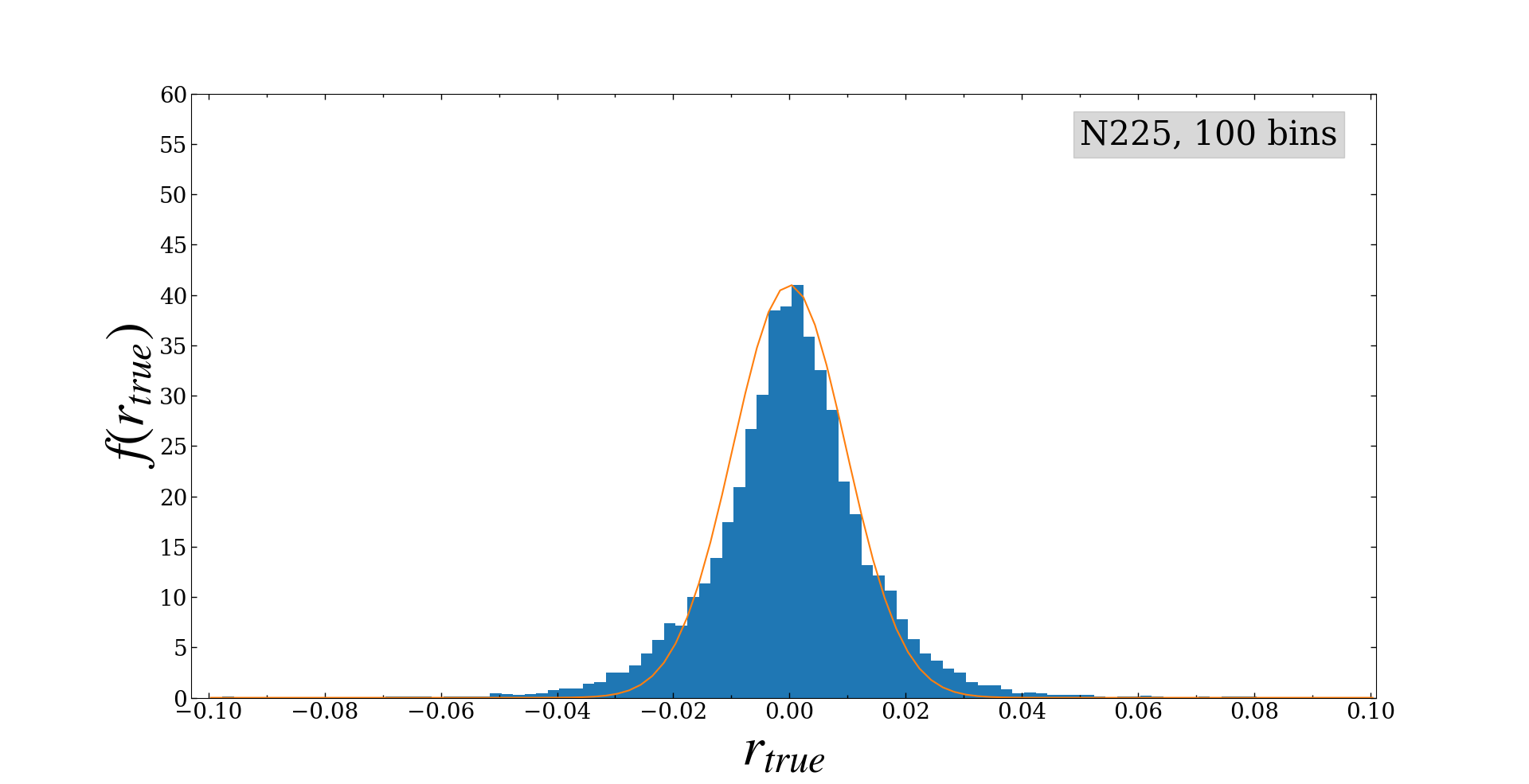}}\hspace{0mm}\\
	\subfloat[The true distribution histogram and raw normal distribution density function curve of N225 daily returns when the number of bins is 150. The parameters of the hypothetical raw normal distribution are $\mu=4.0\times10^{-4}$ and $\sigma=0.010$.]{
		\includegraphics[scale=0.15]{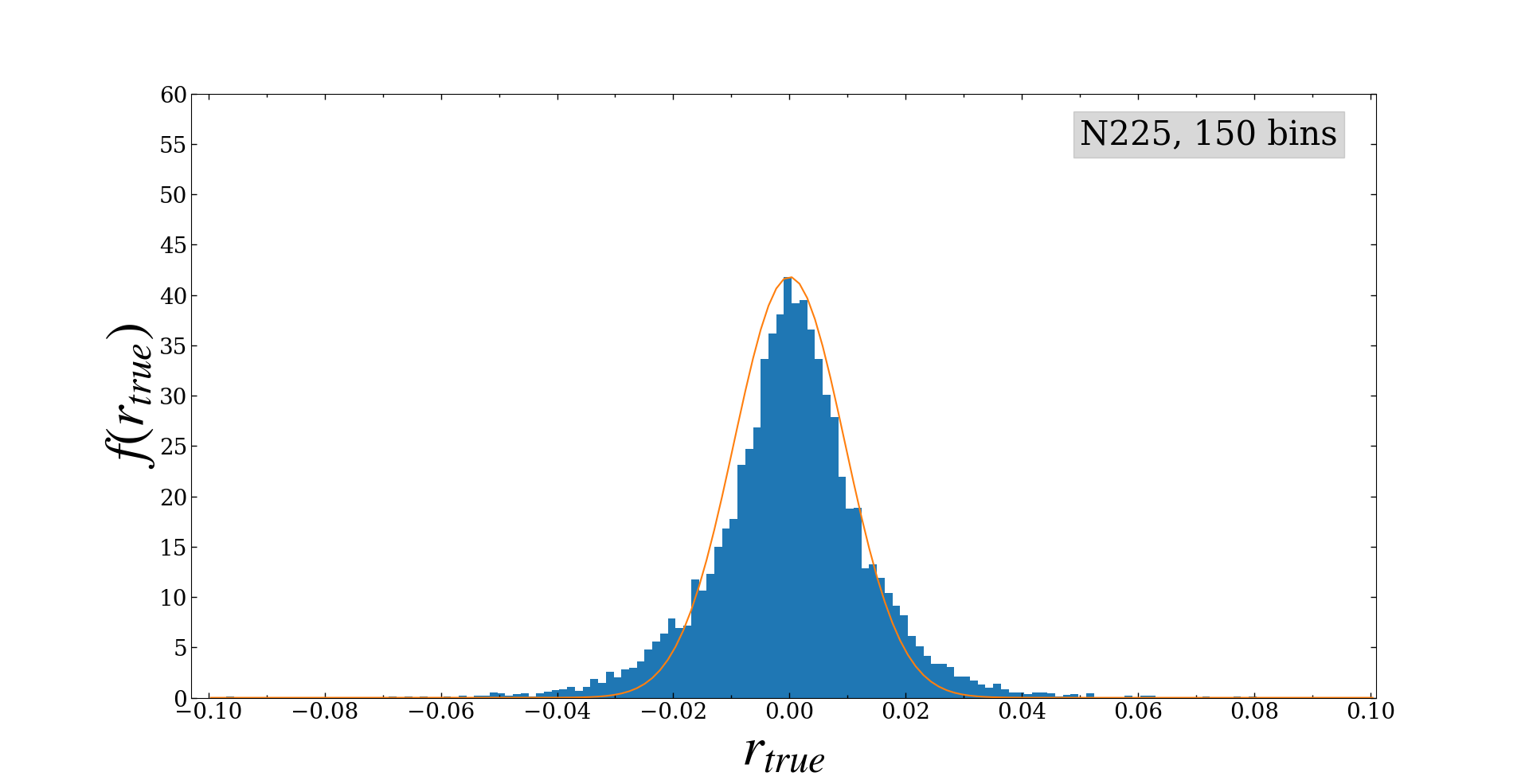}}\hspace{0mm}\\
	\subfloat[The true distribution histogram and raw normal distribution density function curve of N225 daily returns when the number of bins is 200. The parameters of the hypothetical raw normal distribution are $\mu=4.0\times10^{-4}$ and $\sigma=0.010$.]{
		\includegraphics[scale=0.15]{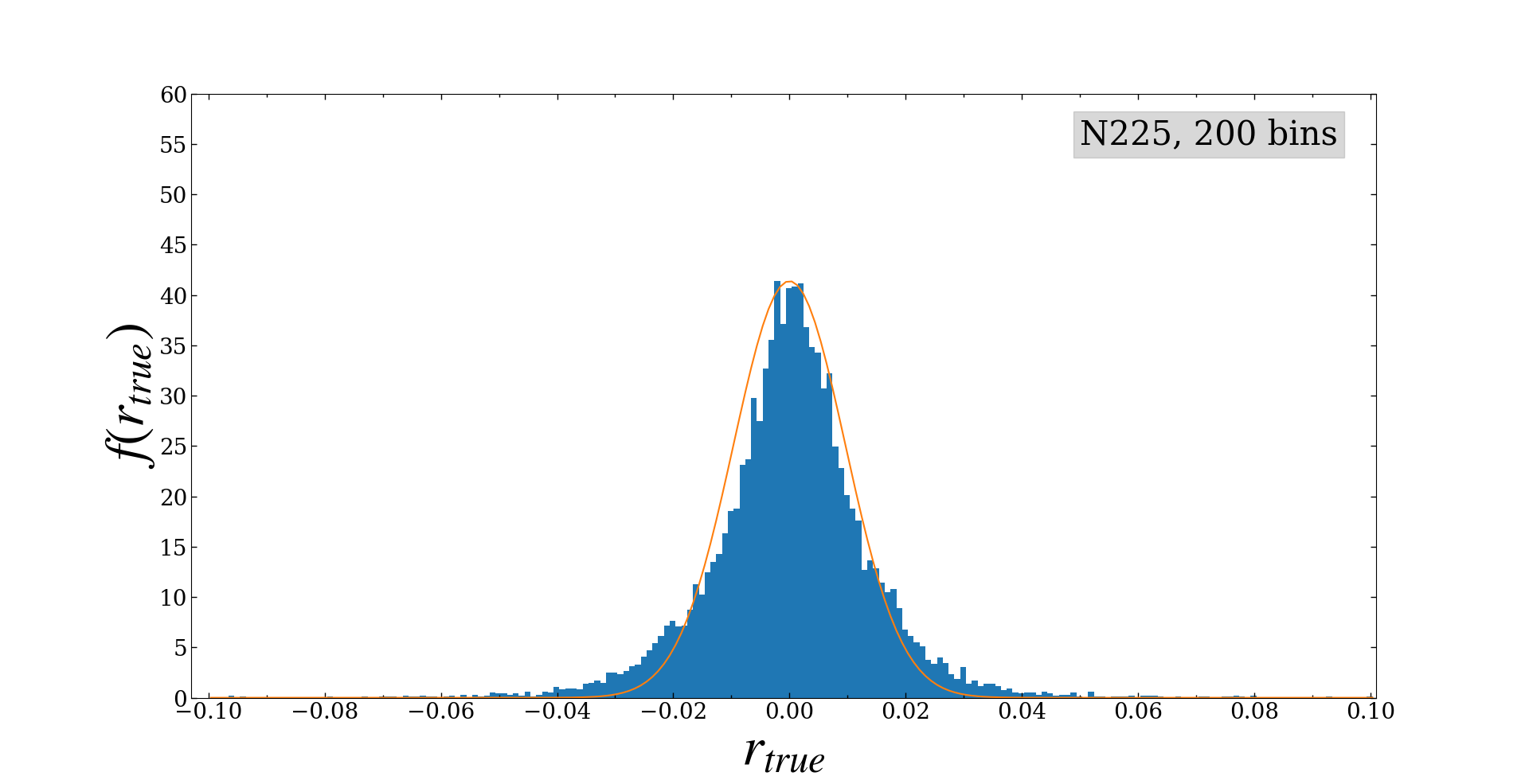}}
	\caption{The true distribution histograms and the hypothetical raw normal distribution density function curves of N225 daily returns when taking different numbers of bins.}
	\label{N225_d}
\end{figure}

Fig.~\ref{CSI_d}, Fig.~\ref{SPX_d} and Fig.~\ref{N225_d} respectively display the true distribution histograms and the hypothetical raw normal distribution density function curves of CSI 300, SPX and N225, when the numbers of the bins are 100, 150 and 200. 
The values of $\sigma$ in the hypothetical raw normal distribution density functions are calculated in the cases of the numbers of bins are taken as different quantities. 
We find the number of bins within our selected range have no significant impact on the value of $\sigma$, the values of $\sigma$ in the hypothetical raw normal distribution density function are around 0.01 for CSI 300, around 0.007 for SPX, and around 0.01 for N225.

From the distribution histograms in Fig.~\ref{CSI_d}, Fig.~\ref{SPX_d} and Fig.~\ref{N225_d}, we can see that a value of 150 for the number of bins is the most appropriate while ensuring that the histograms are smooth and the width of each bin is as narrow as possible. 
When the number of bin is 200, burrs appear on the distribution histograms.

For the minute level stock returns, we also find that the value 150 for the number of bins is the most suitable. 
The analysis is similar to the above and will not be repeated here.

\subsection{Results and Discussion}\label{sec_dis}
\begin{figure}[!hbtp]
	\centering
	\includegraphics[scale=0.18]{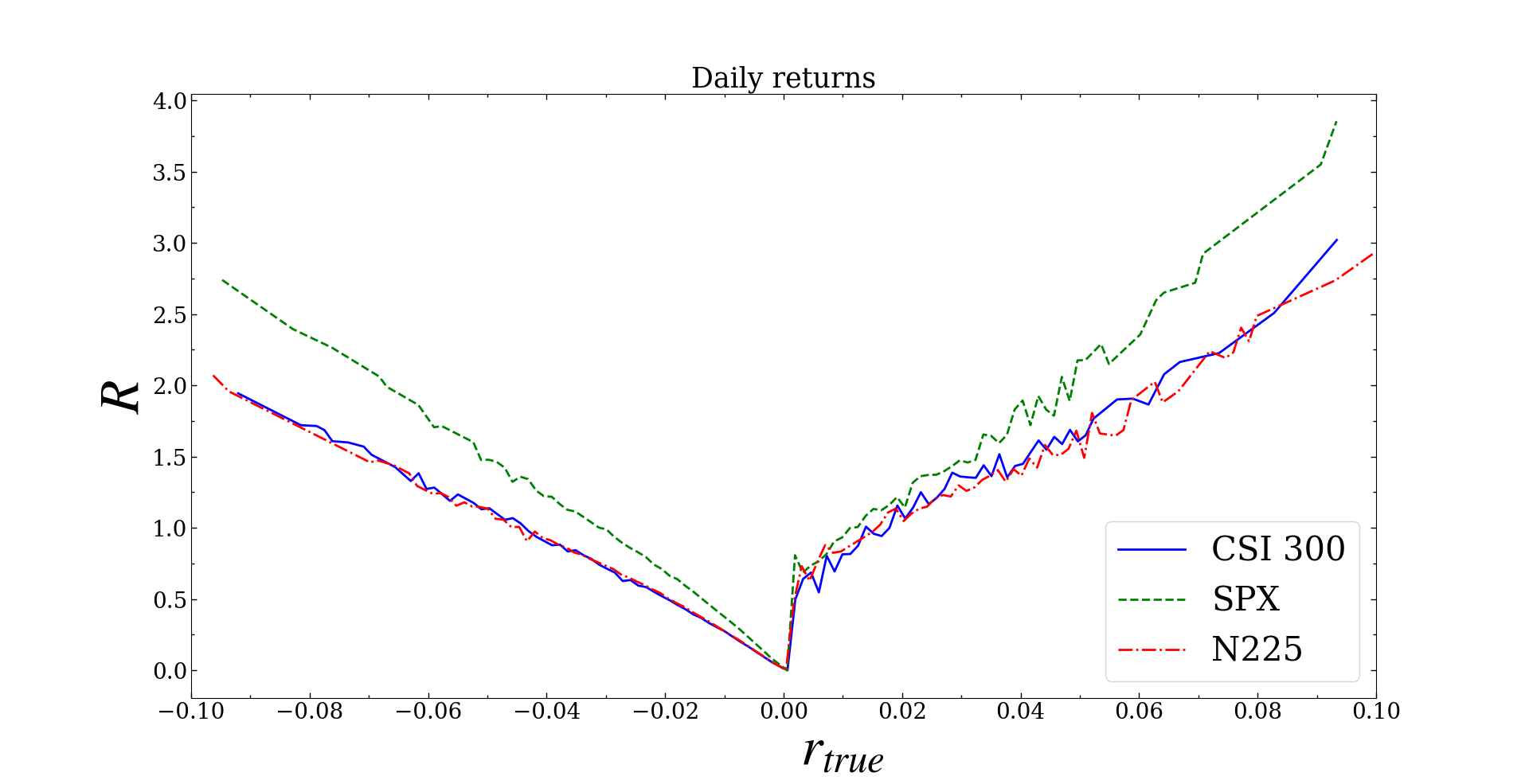}
	\caption{Reaction function curves of daily returns of CSI 300, SPX and N225.}
	\label{R_day}
\end{figure}
Fig.~\ref{R_day} shows the reaction function curves calculated by daily stock returns of CSI 300, SPX, and N225.
From the figure, we can see that true stock returns are the results of financial markets underreact to events or information when the absolute value $|r^{true}|$ approximately below 4\%, and true stock returns are the results of financial markets overacting to events or information when $|r^{true}|$ is approximately above 4\%. 
From these curves, it can be seen that the financial markets' reactions to positive information or events is slightly stronger than that to negative information or events, which results in a skewed distribution of returns.
For the differences in these three reaction functions curves, the reaction functions of CSI 300 and N225 are very close, while the reaction function curve of SPX is relatively steep, whether it is the positive or negative return part.

\begin{figure*}[!htbp]
	\centering
	\subfloat[Reaction function curves of 1-minute returns of CSI 300, SPX and N225.]{
		\includegraphics[scale=0.12]{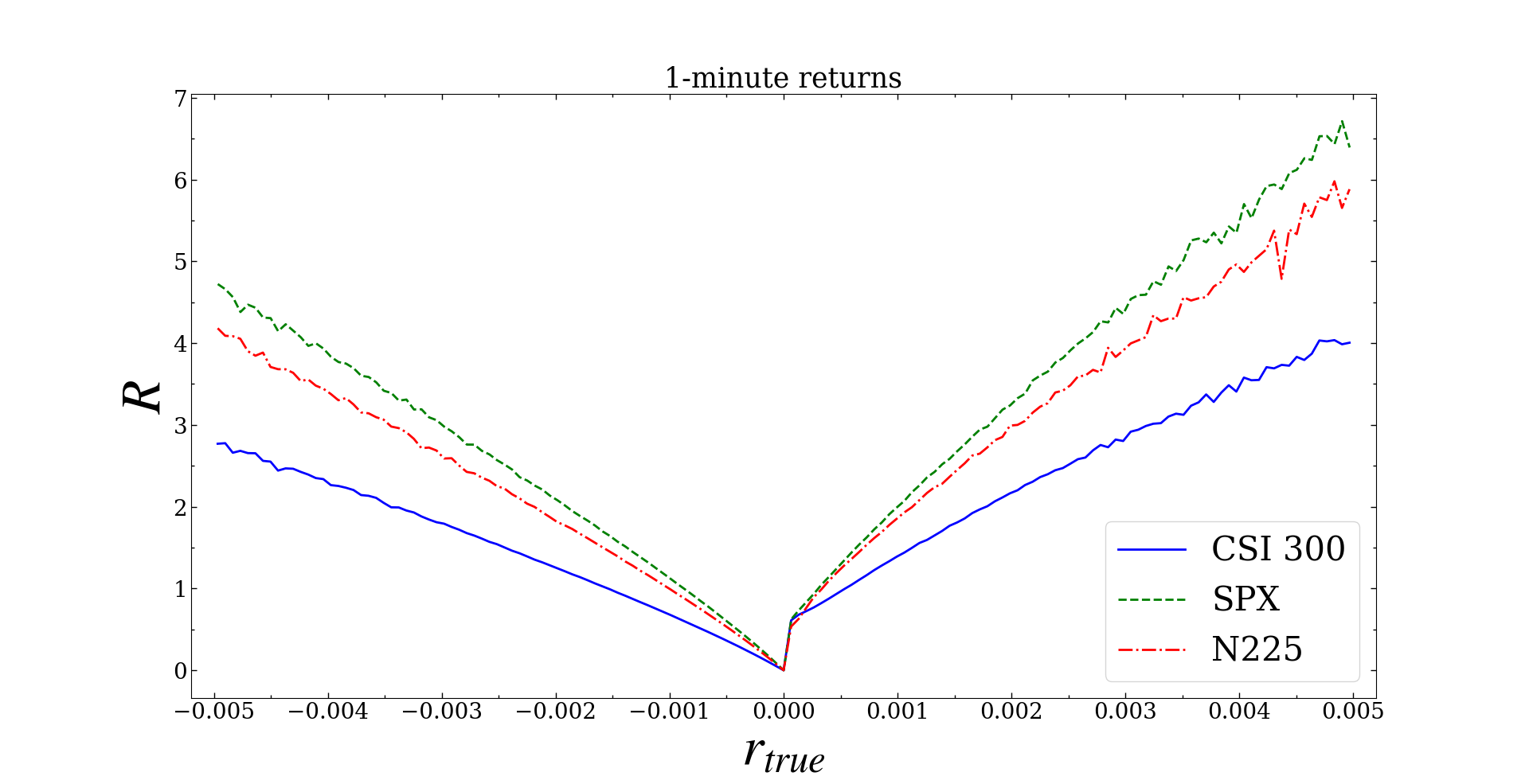}}\hspace{-2mm}
	\subfloat[Reaction function curves of 5-minute returns of CSI 300, SPX and N225.]{
		\includegraphics[scale=0.12]{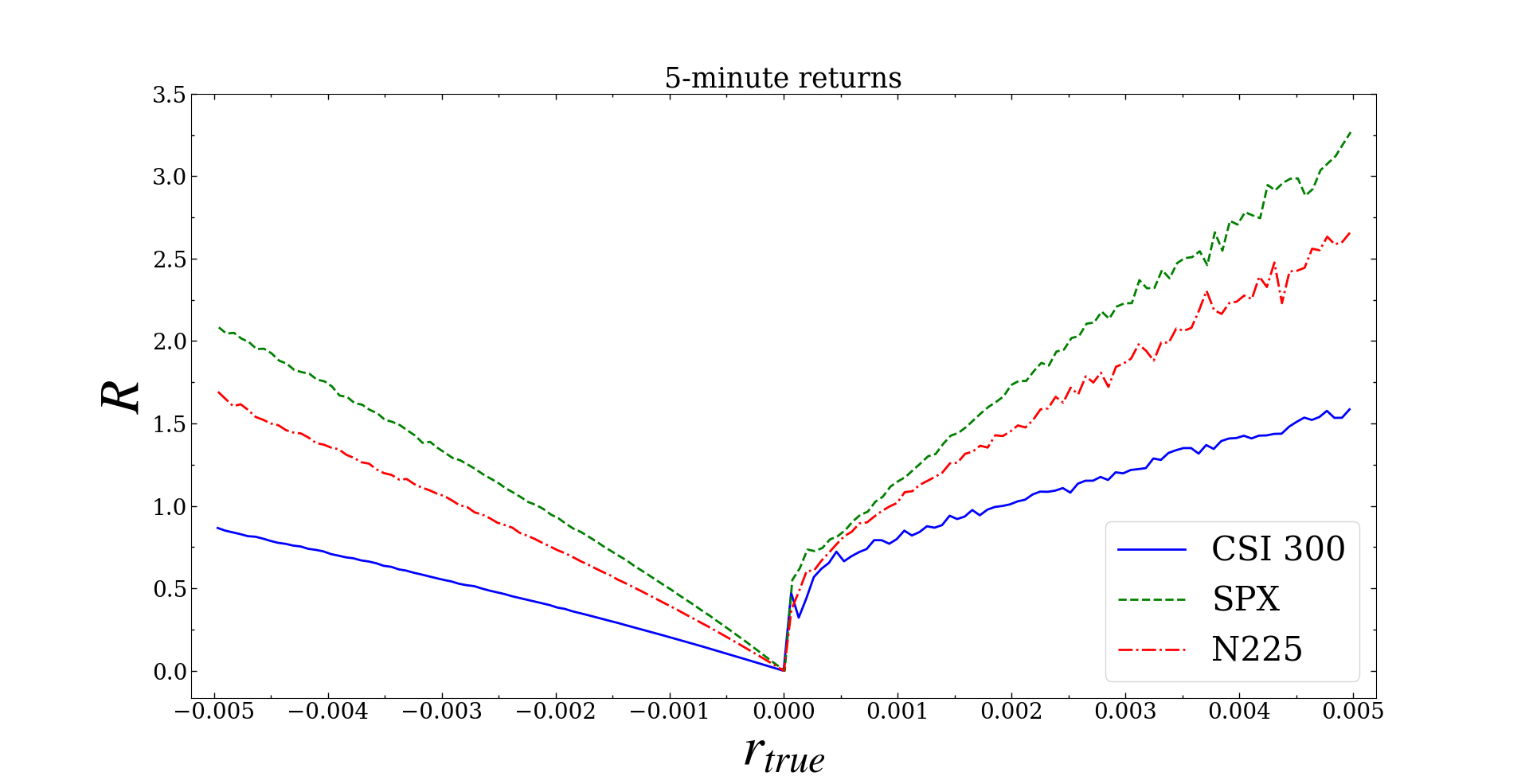}}\hspace{-2mm}
	\subfloat[Reaction function curves of 15-minute returns of CSI 300, SPX and N225.]{
		\includegraphics[scale=0.12]{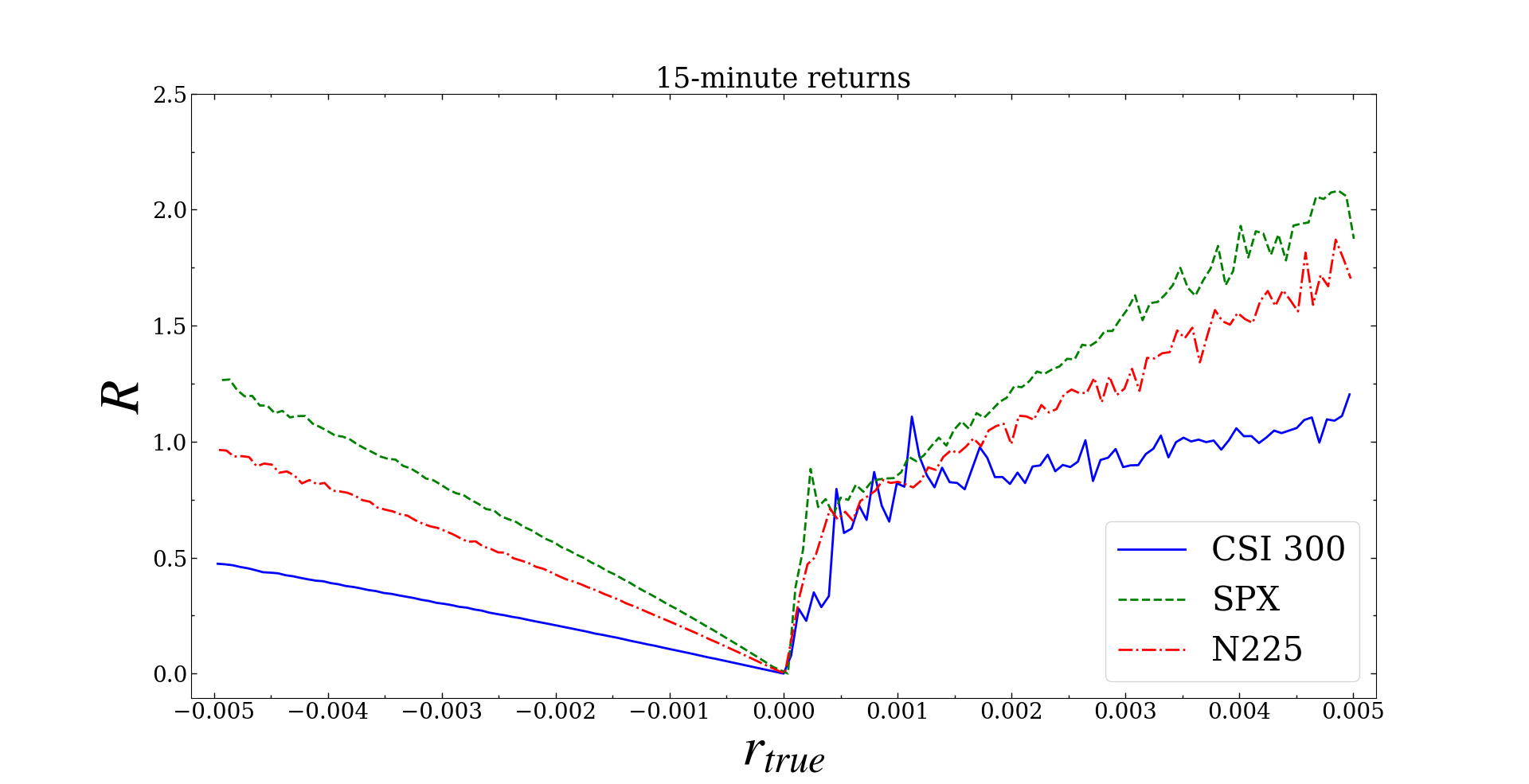}}
	\caption{Reaction function curves of 1-minute, 5-minute, 15-minute returns of CSI 300, SPX and N225.}
	\label{R_min}
\end{figure*} 
Furthermore, we studied the reaction function curves calculated by minute level returns.
Fig.~\ref{R_min} shows the reaction function curves of 1-minute returns, 5-minute returns and 15-minute returns.
From the figure, it can be seen that the characteristics of the reaction function calculated using minute level returns are roughly the same as those reflected when using daily data. 
However, it looks that the market has stronger reactions to events or information from the perspective of minute level scales, which is reflected in the shapes of the reaction functions.
Through the calculation results, we can see that the financial markets generally react more to events that occur less frequently but have greater impacts on financial markets.
The main differences reflected by the three major indices are the reaction function curves of the S\&P 500 are the steepest, followed by N225, and the reaction function curves of the CSI 300 are the smoothest.

In the results about daily returns, we noticed that, the reaction functions of CSI 300 and N225 are very close, while the reaction function curve of SPX is relatively steep, whether the returns are positive or negative.
This may be because the US stock market is more international and can attract more investors, leading to stronger responses of financial markets to information or events.

In the results about 1-minute, 5-minute and 15-minute returns, the reaction function curves of the S\&P 500 are also the steepest, followed by N225, and the reaction function curves of the CSI 300 are relatively flat.
The results are related to the trading mechanisms and the activity level of financial markets. 
As mentioned above, US stocks are the most active, and the stock market trading mechanisms in the United States are the most flexible, followed by the Japanese stock market, while stocks in the Chinese stock market do not allow intraday trading, which limits the market's responses to various financial market events or information.

Why is the reaction function curves calculated through minute level steeper than using daily data, and the shorter the time interval between returns, the steeper the reaction function curves?
We believe that the daily returns mainly digest and react to overnight information, while the influencing factors of minute level returns are mainly the previous returns (regarding the previous returns as events or information, especially the recent returns), which is reflected in the correlations existed in the returns series \cite{liu1999statistical, bogachev2007effect, wang2011quantifying, shi2021clustering}. 
Regarding these correlations, the shorter the time interval between returns, the higher the correlations.
And from the perspective of behavioral finance, this correlations are related to momentum effect existed in financial markets \cite{lee2000price, shi2017time}.
Owing to the reasons above, the reaction function curves calculated through minute level returns are steeper than that calculated through daily data, and the shorter the time interval, the steeper they display.
~\\
\section{Conclusion}\label{sec_con}
In summary, based on the assumption that the stock returns distributions exhibit high peaks and fat tails is due to the overreaction or underreaction of financial markets to comprehensive influencing factors, we propose the concept of reaction function, construct the related distribution model and provide a calculation process to calculate the reaction function through real returns in financial markets.
Utilizing the calculation process, we calculated and analyzed the reaction functions corresponding to the daily returns, 15-minute returns, 5-minute returns, and 1-minute returns of CSI 300, SPX, and N225.

Our main conclusion is that the reactions of financial markets to relatively important events or information are stronger than that of mild events or information, and the reactions to positive information or events are slightly stronger than that to negative information or events, which results in a positive bias in the distribution of returns.
For daily returns, there is little difference in the reaction function curves of the three major stock indices, because the daily returns mainly reflect overnight information, there is not much difference among the three major indices in this regard, although the reaction function curve of SPX, which is the most active and flexible trading mechanism, appears slightly steeper.
For minute level returns, the market trading mechanism and activity level have an impact on the steepness of the reaction function curves. 
The higher the market activity level, the steeper the reaction function curves, so the reaction function curves of SPX
is the steepest, followed by N225, and CSI 300 which does not allow intraday trading is relatively flat.
Moreover, on the minute level reaction function curves, we find that the smaller the time scale, the steeper the reaction function.
This is because the shorter the time interval, the stronger the correlation in the returns series.
In addition, it must be pointed out that the reaction function represents the average effect, because the distribution of returns is a statistical result formed by a large number of samples.

Our research can help to explain the characteristics of high peaks, heavy tails and skewnesses in returns distributions, and to analyze the reaction mechanisms of different stock price indices, stocks, and any other securities to events or information. 
In addition, Our work effectively links the statistical characteristics of stock returns with the perspective of behavioral finance and proposes a quantitative model, thus providing valuable insights for solving behavioral finance problems through quantitative methods.
Finally, our research may be applied to quantitative investment or risk management.
For quantitative investment strategies based on event analysis, our research indicates that financial markets usually amplify the impact of important events, so this effect can be utilized to design investment strategies for profits.
For risk management, this study cautions against underestimating the rare and significant events.
~\\
\begin{acknowledgments}
We wish to acknowledge the support of PKU-WUHAN Institute for Artificial Intelligence for this research.
\end{acknowledgments}

\bibliography{references}

\end{document}